\documentclass[journal, 12pt, onecolumn, draftclsnofoot]{IEEEtran}
\usepackage{authblk}
\usepackage{cuted}

\usepackage{url,cite}
\usepackage{amssymb,mathtools}

\usepackage{graphicx}
\usepackage{multirow}

\usepackage{graphicx}         
\usepackage{booktabs}         

\usepackage{arydshln}
\usepackage{bm}
\usepackage{comment}
\includecomment{mycomment}
\newcommand\undermat[2]{%
  \makebox[0pt][l]{$\smash{\underbrace{\phantom{%
    \begin{matrix}#2\end{matrix}}}_{\text{$#1$}}}$}#2}
\newcommand\overmat[2]{%
  \makebox[0pt][l]{$\smash{\overbrace{\phantom{%
    \begin{matrix}#2\end{matrix}}}^{\text{$#1$}}}$}#2}

\usepackage{braket}

\usepackage{enumerate}

\usepackage{tikz}
\usetikzlibrary{tikzmark, shapes, fit,backgrounds}
\usetikzlibrary{matrix,decorations.pathreplacing,angles,quotes}
\usetikzlibrary{patterns, calc, positioning}
\usetikzlibrary{arrows,arrows.meta}

\tikzstyle{line} = [draw, -latex]

\newtheorem{definition}{Definition}

\newtheorem{theorem}{Theorem}

\newtheorem{remark}{Remark}

\DeclarePairedDelimiterX\symp[2]{\langle}{\rangle_\mathbb{S}}{#1 , #2}
\DeclarePairedDelimiterX\ipp[2]{\langle}{\rangle}{#1 , #2}

\newcommand{\cC}{\mathcal{C}}

\newcommand{\bA}{\mathbf{A}}

\newcommand{\bG}{\mathbf{G}}
\newcommand{\bH}{\mathbf{H}}
\newcommand{\bI}{\mathbf{I}}
\newcommand{\bJ}{\mathbf{J}}

\newcommand{\bM}{\mathbf{M}}

\newcommand{\bP}{\mathbf{P}}
\newcommand{\bQ}{\mathbf{Q}}

\newcommand{\be}{\mathbf{e}}

\newcommand{\bv}{\mathbf{v}}

\newcommand{\bx}{\mathbf{x}}
\newcommand{\by}{\mathbf{y}}

\def\ppsmatrix#1{\begin{psmallmatrix}#1\end{psmallmatrix}} 
\def\rowspan#1{\langle #1 \rangle_{\mathsf{row}}}
\def\colspan#1{\langle #1 \rangle_{\mathsf{col}}}

\def\bzero{\mathbf{0}}

\def\F{\mathbb{F}}

\def\Fq{\F_{q}}

\def\i{\imath}								

\def\ppsmatrix#1{\begin{psmallmatrix}#1\end{psmallmatrix}}

\begin{document}
\title{\Large Quantum Cross Subspace Alignment Codes via  the $N$-sum Box Abstraction}
\author{\normalsize Yuxiang Lu and Syed A. Jafar}
\affil{\small Center for Pervasive Communications and Computing (CPCC), UC Irvine\\
Email: \{yuxiang.lu, syed\}@uci.edu}
\date{}

\maketitle

\begin{abstract}
Cross-subspace alignment (CSA) codes are used in various private information retrieval (PIR) schemes (e.g., with secure storage) and in secure distributed batch matrix multiplication (SDBMM). Using a recently developed $N$-sum box abstraction of a quantum multiple-access channel (QMAC), we translate CSA schemes over classical multiple-access channels into efficient quantum CSA schemes over a QMAC, achieving maximal superdense coding gain. Because of the $N$-sum box abstraction, the underlying problem of coding to exploit quantum entanglements for CSA schemes, becomes conceptually equivalent to that of designing a channel matrix for a MIMO MAC subject to given structural constraints imposed by the $N$-sum box abstraction, such that the resulting MIMO MAC is able to implement the functionality of a CSA scheme (encoding/decoding) \emph{over-the-air}. Applications include Quantum PIR with secure and MDS-coded storage, as well as Quantum SDBMM.
\end{abstract}

\section{Introduction}
A central challenge in  sending \emph{classical} information over \emph{quantum} communication \emph{networks} is to find coding schemes that exploit quantum entanglements  to improve communication efficiency. The challenge is  compounded when  the goal is \emph{secure/private computation}. Particularly relevant for our purpose are the recent works in \cite{song2019capacity, QTPIR, QMDSTPIR} that develop  coding schemes for  quantum private information retrieval (QPIR), i.e., private  retrieval of \emph{classical} information over an ideal quantum multiple-access channel (QMAC), where the communication takes place via qudits instead of classical dits.\footnote{A dit is a classical $q$-ary symbol, e.g., a symbol in a finite field $\mathbb{F}_q$ when $q$ is a power of a prime, whereas a qudit is a $q$-dimensional quantum system.}

Recall that the goal in PIR \cite{PIRfirstjournal} is for a user to efficiently \cite{Sun_Jafar_PIR} retrieve a desired message from a set of servers where multiple messages are stored, without revealing which message is desired. To do so, the user sends queries to the servers that reveal nothing about the desired message index. The servers respond with their answers, and from the answers the user retrieves the desired message. The communication from the servers to the user constitutes a multiple-access channel (MAC). The multiple-access aspect is trivial  in  classical PIR  because each server has a separate  channel to the user over which classical symbols (dits) can be transmitted. In QPIR \cite{song2019capacity, QTPIR, QMDSTPIR}, however, instead of dits, the servers send qudits to the user, into which they encode their classical answers through local quantum operations. While the servers still send their qudits through \emph{separate} quantum channels, what makes the QMAC non-trivial is that the qudits available to the servers are entangled with each other. The entanglements are designed \emph{a priori}, i.e., independent of messages and queries,  each server can only manipulate its own qudits locally, and the servers cannot directly communicate with each other (except for collusions that may be tolerated). Yet the QMAC setting is non-trivial because the local operations (quantum gates) can be carefully chosen by the servers based on their classical answers, to create a new \emph{entangled} state that can more efficiently convey the desired information to the user through quantum measurement. Indeed, coding schemes in \cite{song2019capacity, QTPIR, QMDSTPIR} exploit quantum entanglements to reduce the user's total download cost by upto a factor of $2$ (maximal superdense coding gain).\footnote{Superdense coding refers to the capability of each  qudit to deliver more than $1$ dit (maximally $2$ dits) of classical information. The key is to exploit entanglements (without which qudits are worth no more than dits \cite{Holevo_Bounds}). This is challenging to accomplish when the data is distributed, as in the QMAC.} In addition to  QPIR  \cite{song2019capacity}, these prior works also explore variants that allow upto $T$ colluding servers (QTPIR) \cite{QTPIR} and MDS coded storage (QMDSTPIR) \cite{QMDSTPIR}, developing quantum  schemes for each setting that are  counterparts of the corresponding classical schemes developed in  \cite{Sun_Jafar_TPIR, Banawan_MDSPIR,Sun_Jafar_MDSTPIR,Tajeddine_MDSTPIR}.

As noted in \cite{PIR_tutorial}, coding schemes for PIR  often find applications beyond PIR. Particularly noteworthy are \emph{cross subspace alignment} (CSA) codes which originated in the studies of PIR with $X$-secure  (XSTPIR) \cite{Jia_Sun_Jafar_XSTPIR} and MDS coded (MDSXSTPIR) storage \cite{Jia_Jafar_MDSXSTPIR},  later found use in coded computing, and were applied, e.g., to secure distributed batch matrix multiplication (SDBMM) in \cite{Jia_Jafar_SDMM, Chen_Jia_Wang_Jafar_NGCSA}. If classical CSA codes can be translated into quantum coding schemes for the QMAC, it would open the door  not only to code designs for broader QPIR settings (e.g., QXSTPIR and QMDSXSTPIR), but also to  quantum secure distributed batch matrix multiplication (QSDBMM). This is the goal that we pursue in this work.

While the goal ostensibly involves quantum coding, the  approach in this work requires little quantum-expertise because of the $N$-sum box abstraction  \cite{Allaix_N_sum_box}. Conceptually, the technical problem that we need to solve can be interpreted as designing the channel matrix for a wireless MIMO MAC to efficiently implement the CSA encoder/decoder structure \emph{over-the-air}. The main distinctions from the multiple antenna wireless setting are 1) the channel matrix is over a finite field ($\mathbb{F}_q$) rather than complex numbers, and 2) instead of  being generated randomly by nature, the channel matrix can be freely designed for the desired CSA code functionality \emph{as long as} it is \emph{strongly self-orthogonal} (SSO).\footnote{\label{fn:sso}A matrix ${\bf G}\in\mathbb{F}_q^{2N\times N}$ is strongly self-orthogonal (SSO) iff $\mbox{rk}({\bf G})=N$ and ${\bf G}^T{\bf J}{\bf G}={\bf 0}$, where ${\bf J}\triangleq \ppsmatrix{{\bf 0}&-{\bf I}_N\\ {\bf I}_N&{\bf 0}}$.} This  is because  the $N$-sum box abstraction \cite{Allaix_N_sum_box} guarantees that a stabilizer based quantum coding protocol exists that achieves the same functionality as a classical MIMO MAC setting for any SSO channel matrix, with the quantum communication cost of $N$ qudits.  The CSA scheme and the $N$-sum box abstraction are explained next.

\subsection{Cross Subspace Alignment (CSA) Code}
Conceptually, the setting for a CSA coding scheme (over a finite field $\mathbb{F}_q$) is the following.  We have $N$ distributed servers. The servers locally compute their answers $A_n, n\in[N]$ to a user's query. Each answer $A_n$ is a linear combination  of $L$ symbols that are desired by the user, say $\delta_1, \delta_2, \cdots, \delta_L$, and $N-L$ symbols of undesired information (interference), say $\nu_1, \nu_2, \cdots, \nu_{N-L}$. The linear combinations have a Cauchy-Vandermonde structure that is the defining characteristic of CSA schemes, such that the desired terms appear along the dimensions corresponding to the Cauchy terms, while the interference  appears (aligned) along the dimensions corresponding to the Vandermonde terms. The $i^{th}$ instance of a CSA scheme is represented as,
\begin{align*}
&\scalebox{0.75}
{$\underbrace{\begin{bmatrix}
A_1(i)\\ \vdots \\ A_N(i)
\end{bmatrix}}_{\bm{A}(i)}
=$}
\scalebox{0.75}
{$\underbrace{\left[\begin{array}{ccc|cccc}
\frac{1}{f_1-\alpha_1}&\cdots&\frac{1}{f_L-\alpha_1}&1&\alpha_1&\cdots&\alpha_1^{N-L-1}\\
\frac{1}{f_1-\alpha_2}&\cdots&\frac{1}{f_L-\alpha_2}&1&\alpha_2&\cdots&\alpha_2^{N-L-1}\\
\vdots&\vdots&\vdots&\vdots&\vdots&\vdots&\vdots\\
\frac{1}{f_1-\alpha_N}&\cdots&\frac{1}{f_L-\alpha_N}&1&\alpha_N&\cdots&\alpha_N^{N-L-1}\\
\end{array}\right]}_{\mbox{CSA}^q_{N,L}(\bm{\alpha},\bm{f})}
$}
\scalebox{0.75}
{$\underbrace{\begin{bmatrix}
\delta_1(i)\\
:\\
\delta_L(i)\\
\nu_1(i)\\
:\\
\nu_{\mbox{\tiny $N-L$}}(i)
\end{bmatrix}}_{\bm{X}_{\delta\nu}(i)}
$}
\end{align*}
The CSA scheme requires  that all the $\alpha_i, f_j$ are distinct elements in $\mathbb{F}_q$ (needs $q\geq N+L$), which guarantees that the $N\times N$ matrix $\mbox{CSA}^q_{N,L}(\bm{\alpha},\bm{f})$ is invertible. 
After downloading $A_n$ from each Server $n, n\in[N]$, the user is able to recover the desired symbols $\delta_1(i), \cdots, \delta_L(i)$ by inverting $\mbox{CSA}^q_{N,L}(\bm{\alpha},\bm{f})$. Thus, each instance of the CSA scheme allows the user to retrieve $L$ desired symbols at a cost of $N$ downloaded symbols. The rate of the scheme, defined as the number of desired symbols  recovered per downloaded symbol,  is  $L/N$. The reciprocal,  $N/L$, is the download cost per desired symbol.

\begin{remark}\label{rem:redundant} A noteworthy aspect of  CSA schemes is that the number of servers can be reduced, i.e., the CSA scheme can be applied to $N'<N$ servers, with a corresponding reduction in the number of desired symbols $L'<L$, as long as the dimension of interference is preserved, i.e., $N-L=N'-L'$. In the classical setting, this flexibility is not useful as it leads to a strictly higher download cost, i.e., $N'/L' = (N-L)/L' + 1>(N-L)/L+1=N/L$. In the quantum setting, however, this will lead to  a useful simplification.
\end{remark}

\subsection{$N$-Sum Box Abstraction as a MIMO MAC}
\begin{figure}[h]
\centering
\begin{tikzpicture}
\node (Tx1) [draw, rectangle, rounded corners, thick, minimum width=0.52cm, minimum height=0.7cm] {\footnotesize Tx1};
\node (x1) [draw, thick, circle, inner sep = 0.05cm, above right=-0.25cm and 0.5cm of Tx1] {};
\node (x4) [draw, thick, circle, inner sep = 0.05cm, below right=-0.25cm and 0.5cm of Tx1] {};
\draw [thick] (x1.west)--(Tx1.east|-x1.west) node [midway, above=-0.1cm]{\footnotesize $x_1$};
\draw [thick] (x4.west)--(Tx1.east|-x4.west) node [midway, above=-0.1cm]{\footnotesize $x_{4}$};
\begin{scope}[yshift=-1cm]
\node (Tx2) [draw, rectangle, rounded corners, thick, minimum width=0.52cm, minimum height=0.7cm] {\footnotesize Tx2};
\node (x2) [draw, thick, circle, inner sep = 0.05cm, above right=-0.25cm and 0.5cm of Tx2] {};
\node (x5) [draw, thick, circle, inner sep = 0.05cm, below right=-0.25cm and 0.5cm of Tx2] {};
\draw [thick] (x2.west)--(Tx2.east|-x2.west) node [midway, above=-0.1cm]{\footnotesize $x_2$};
\draw [thick] (x5.west)--(Tx2.east|-x5.west) node [midway, above=-0.1cm]{\footnotesize $x_{5}$};
\end{scope}
\begin{scope}[yshift=-2cm]
\node (Tx3) [draw, rectangle, rounded corners, thick, minimum width=0.52cm, minimum height=0.7cm] {\footnotesize Tx3};
\node (x3) [draw, thick, circle, inner sep = 0.05cm, above right=-0.25cm and 0.5cm of Tx3] {};
\node (x6) [draw, thick, circle, inner sep = 0.05cm, below right=-0.25cm and 0.5cm of Tx3] {};
\draw [thick] (x3.west)--(Tx3.east|-x3.west) node [midway, above=-0.1cm]{\footnotesize $x_3$};
\draw [thick] (x6.west)--(Tx3.east|-x6.west) node [midway, above=-0.1cm]{\footnotesize $x_{6}$};
\end{scope}

\node (Rx) [draw, rectangle, rounded corners, thick, minimum width=0.5cm, minimum height=1.5cm, right=2.5cm of Tx2] {\footnotesize \rotatebox{90}{Receiver}};
\node (y2) [draw, thick, circle, inner sep = 0.05cm,   left= 0.5cm of Rx] {};
\node (y1) [draw, thick, circle, inner sep = 0.05cm,  above= 0.2cm of y2] {};
\node (y3) [draw, thick, circle, inner sep = 0.05cm,  below= 0.2cm of y2] {};

\draw [thick] (y1.east)--(Rx.west|-y1.east) node [midway, above=-0.1cm]{\footnotesize $y_1$};
\draw [thick] (y2.east)--(Rx.west|-y2.east) node [midway, above=-0.1cm]{\footnotesize $y_2$};
\draw [thick] (y3.east)--(Rx.west|-y3.east) node [midway, above=-0.1cm]{\footnotesize $y_3$};

\foreach \i in {1,2,...,6},
	\foreach \j in {1,2,3},
		\draw [->, help lines] (x\i)--(y\j);
		
\node [right=0.61cm of Rx, align=left]{\scalebox{0.75}{$\begin{bmatrix}y_1\\y_2\\y_3\end{bmatrix}=\begin{bmatrix}M_{1,1}&\cdots&M_{1,6}\\ \vdots&\vdots&\vdots\\ M_{3,1}&\cdots&M_{3,6}\end{bmatrix}\begin{bmatrix}x_1\\x_2\\\vdots\\x_6\end{bmatrix}$}\\[0.1cm] \footnotesize ${\bf y}\in\mathbb{F}_q^{3\times 1}, {\bf x}\in\mathbb{F}_q^{6\times 1}, {\bf M}\in\mathbb{F}_q^{3\times 6}$\\[0.1cm]\footnotesize ${\bf M}^T{\bf J}{\bf M}={\bf 0}$, $\mbox{rk}({\bf M})=3$ (SSO)\\[0.1cm] \footnotesize {\color{black}[Quantum Comm. Cost: $3$ qudits]}};
\end{tikzpicture}
\caption{The $N$-sum box \cite{Allaix_N_sum_box} is illustrated as a MIMO MAC  for $N=3$.}\label{fig:MIMOMAC}
\end{figure}
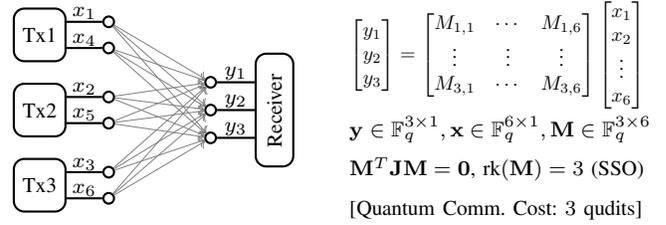
Let $\mathbb{F}_q$ be a finite field. 
An $N$-sum box \cite{Allaix_N_sum_box} is a MIMO MAC  setting described as,
    $\mathbf{y} = \mathbf{M}\mathbf{x}$,
where $\mathbf{x} = [x_1 ~~ x_2 ~~\cdots~~ x_{2N}]^\top \in \mathbb{F}_q^{2N\times 1}$ is the input, 
$\mathbf{M} \in \mathbb{F}_{q}^{N\times 2N}$ is the `\emph{channel}' matrix and $\mathbf{y} \in  \mathbb{F}_{q}^{N\times 1}$ is the output. As shown in \cite{Allaix_N_sum_box}, if ${\bf M}^\top$ is strongly self-orthogonal (SSO), then this $N$-sum box  functionality is achievable in a QMAC with $N$ servers, with $N$ entangled $q$-dimensional qudits distributed among the servers, such that the $n^{th}$ server, who has the $n^{th}$ qudit, controls the classical inputs $x_n, x_{N+n}$ in the MIMO MAC, $\forall n\in[N]$. An example is illustrated in Fig. \ref{fig:MIMOMAC}. For any feasible channel matrix ${\bf M}$, it is shown in \cite{Allaix_N_sum_box} that there exists an initial entangled state of the $N$ qudits, such that each transmitter can  encode its classical inputs $x_n, x_{n+N}$ into its qudit through local quantum operations ($X$, $Z$ gates), and there exists a quantum measurement such that once all $N$ $q$-dimensional qudits are sent to the receiver (quantum communication cost $N$ $q$-dimensional qudits), the receiver obtains the classical value ${\bf y}$ from the measurement. 

An alternative, \emph{equivalent} description of feasible channel matrices ${\bf M}$ in \cite{Allaix_N_sum_box} that is better suited to our purpose is the following. For any $\mathbf{G} \in \mathbb{F}_{q}^{2N \times N}, \mathbf{H} \in \mathbb{F}_{q}^{2N \times N}$ such that ${\bf G}$ is SSO and the square matrix $(\mathbf{G}~~\mathbf{H})$ has full rank $2N$,
there exists an $N$-sum box over $\mathbb{F}_q$ with the channel matrix, 
\begin{align}
    \mathbf{M} = (\mathbf{0}_{N} ~~ \mathbf{I}_{N})(\mathbf{G}~~\mathbf{H})^{-1}.\label{eq:feasible}
\end{align}

\subsection{Overview of Contribution}
Since the MIMO MAC obtained from the $N$-sum box abstraction produces linear combinations of the inputs at the receiver, and the channel matrix ${\bf M}$ can be carefully chosen subject to certain (SSO) constraints, there is the opportunity to design a suitable channel matrix (and therefore an efficient quantum code for the underlying QMAC), that facilitates the CSA encoder/decoder operations, by implementing them `\emph{over-the-air}' as it were (\emph{actually} through quantum-entanglements), as much as possible. As our main contribution we find a feasible channel matrix construction, i.e., a feasible ${\bf M}$, and thereby a Quantum CSA scheme for $N$ servers, that retrieves $\min(1,2r)$ desired classical $q$-ary dits per downloaded $q$-dimensional qudit, if the corresponding classical CSA scheme retrieves $r$ desired dits per downloaded dit. The QCSA scheme  achieves the same rate improvements for QXSTPIR, QMDSXSTPIR and QSDBMM, compared to the classical rates achieved by CSA schemes for XSTPIR, MDSXSTPIR and SDBMM. Furthermore, since QPIR, QTPIR, and QMDSTPIR are special cases of these settings, the rates achieved  in prior works \cite{song2019capacity, QTPIR,QMDSTPIR} can also be obtained as a special case of the QCSA scheme developed in this work.

{\it Notation:} For  integers $a, b$ s.t. $a \leq b$,  define $[a:b]\triangleq\{a,a+1,\cdots,b\}$, and $[b]\triangleq[1:b]$.  $\mbox{Diag}(\bv)$ denotes the diagonal  matrix with entries of $\bv$ along the diagonal. For an $m\times n$ matrix $\mathbf{A}$, $\mathbf{A}(a:b,c:d)$, where  $1 \leq a \leq b \leq m$ and $1 \leq c \leq d \leq n$, denotes the matrix comprised of the entries of $\mathbf{A}$ whose row indices are in $[a:b]$ and column indices are in $[c:d]$. We use $\mathbf{A}(:,c:d)$ to denote the $c^{th}$ to $d^{th}$ columns of $\bA$ and $\mathbf{A}(a:b,:)$ to denote the $a^{th}$ to $b^{th}$ rows.  $\colspan{\bA}$ and $\rowspan{\bA}$ denote the column space and row space of  $\bA$, respectively. $\bA^\top$ is the transpose, and $\mbox{rk}(\bA)$ is the rank of $\bA$.

\section{Definitions}\label{sec:pre}
Let $\mathbb{F}_q$ be a finite field. We need the following definitions.
\vspace{0.2cm}
\begin{definition}{\bf (GRS Code)} Define
\begin{align*}
\mbox{\rm GRS}^q_{n,k}(\bm{\alpha},\bm{u})&\triangleq \scalebox{0.81}{$\begin{bmatrix}
u_1&u_1\alpha_1&u_1\alpha_1^2&\cdots&u_1\alpha_1^{k-1}\\[0.1cm]
u_2&u_2\alpha_2&u_2\alpha_2^2&\cdots&u_2\alpha_2^{k-1}\\[0.1cm]
\vdots&\vdots&\vdots&\vdots&\vdots\\[0.1cm]
u_n&u_n\alpha_n&u_n\alpha_n^2&\cdots&u_n\alpha_n^{k-1}\\[0.1cm]
\end{bmatrix}_{n\times k}$}
\end{align*}
as the generator matrix of an $[n,k]$ Generalized Reed Solomon (GRS) code over $\mathbb{F}_q$, where $\bm{\alpha}=(\alpha_1,\alpha_2,\cdots,\alpha_n)\in\mathbb{F}_q^{1\times n}$, $\bm{u}=(u_1,u_2,\cdots,u_n)\in\mathbb{F}_q^{1\times n}$, $u_i\neq 0$ and $\alpha_i\neq \alpha_j$ for all distinct $i,j\in[n]$. The  corresponding $[n,k]$ GRS code is the column-span of the generator matrix, $\cC=\colspan{\mbox{\rm GRS}^q_{n,k}(\bm{\alpha},\bm{u})}$.
\end{definition}
\vspace{0.2cm}
\begin{definition}{\bf (Dual Code)}
For an $[n,k]$ linear code $\cC\subset \mathbb{F}_q^{n\times 1}$, the dual code $\cC^\perp\triangleq\{{\bf v}\in\mathbb{F}_q^{n\times 1}: \langle\bf{v},\bf{c}\rangle=0, \forall \bf{c}\in \cC\}.$
\end{definition}
The dual code of a GRS code is also a GRS code, e.g., according to the following construction \cite{Macwilliams}.
\vspace{0.2cm}
\begin{definition}{{\bf (Dual GRS Code)}\cite{Macwilliams}}
For $\bm{v}=(v_1,v_2,\cdots,v_n)\in\mathbb{F}_q^{1\times n}$ defined as,
\begin{align}
v_j&=\frac{1}{u_j}\left(\prod_{i\in[n],i\neq j }(\alpha_j-\alpha_i)\right)^{-1}&&\forall j\in[n],\label{def:v}
\end{align}
where $u_j \neq 0$, we have
\begin{align}
\Big(\mbox{\rm GRS}^q_{n,k}(\bm{\alpha},\bm{u})\Big)^\top\mbox{\rm GRS}^q_{n,n-k}(\bm{\alpha},\bm{v})&=\bm{0}_{k\times (n-k)},\label{eq:guarantee}
\end{align}
i.e., $\mbox{\rm GRS}^q_{n,n-k}(\bm{\alpha},\bm{v})$ generates an $[n,n-k]$ code that is the dual of the code generated by $\mbox{\rm GRS}^q_{n,k}(\bm{\alpha},\bm{u})$.
\end{definition}
\begin{definition}{\bf (QCSA Matrix)} \label{def:QCSA} Define the $N\times N$ matrix, $\mbox{\rm QCSA}^q_{N,L}(\bm{\alpha},\bm{\beta},\bm{f})$ as in \eqref{eq:MCV_Blocks},
    \vspace{0.4cm}
    \begin{align}
        &\scalebox{0.80}{$\mbox{\rm QCSA}^q_{N,L}(\bm{\alpha},\bm{\beta},\bm{f})$}
        \triangleq
        \notag\\
        &
        \scalebox{0.80}{$
        \begin{bmatrix}
            \begin{array}{cccc|ccccccccc}
                    \frac{\beta_1}{f_1 - \alpha_1} & \frac{\beta_1}{f_2 - \alpha_1} & \cdots & \frac{\beta_1}{f_L - \alpha_1}
                    & \overmat{\mbox{\rm GRS}^q_{N,\lfloor N/2\rfloor}(\bm{\alpha},\bm{\beta})}{\beta_1 & \beta_1 \alpha_{1} & \cdots &  \beta_1 \alpha_{1}^{\lfloor N/2\rfloor - 1}} & \beta_1 \alpha_{1}^{\lceil N/2\rceil - 1}
                     &\beta_1 \alpha_{1}^{\lceil N/2\rceil} & \cdots & \beta_1 \alpha_{1}^{N-L-1}\\[0.1cm]
                    \frac{\beta_2}{f_1 - \alpha_2} & \frac{\beta_2}{f_2 - \alpha_2} & \cdots & \frac{\beta_2}{f_L - \alpha_2}
                    &\beta_2 & \beta_2 \alpha_{2} & \cdots & \beta_{2} \alpha_{2}^{\lfloor N/2\rfloor - 1} & \beta_2 \alpha_{2}^{\lceil N/2\rceil - 1}
                     &\beta_2 \alpha_{2}^{\lceil N/2\rceil} & \cdots & \beta_2 \alpha_{2}^{N-L-1}\\
                    \vdots & \vdots & \vdots & \vdots&\vdots & \vdots & \vdots&\vdots & \vdots & \vdots\\
                     \undermat{\tilde{\bH}^{\bm{\beta}}(:,1:L)}{\frac{\beta_N}{f_1 - \alpha_N} & \frac{\beta_N}{f_2 - \alpha_N} & \cdots & \frac{\beta_N}{f_L - \alpha_N}}&
                     \undermat{\mbox{\rm GRS}^q_{N,\lceil N/2\rceil}(\bm{\alpha},\bm{\beta})} {\beta_N & \beta_N \alpha_{N} & \cdots & \beta_N \alpha_{N}^{\lfloor N/2\rfloor - 1} & \beta_N \alpha_{N}^{\lceil N/2\rceil - 1}}&
                       \undermat{\tilde{\bH}^{\bm{\beta}}(:,L+1:\lfloor N/2\rfloor)} {\beta_N \alpha_{N}^{\lceil N/2\rceil} & \cdots & \beta_N \alpha_{N}^{N-L-1}}
            \end{array}
            \end{bmatrix}$
        }\label{eq:MCV_Blocks}\\
        \notag
    \end{align}
where $\bm{\alpha}=(\alpha_1,\alpha_2,\cdots,\alpha_N)\in\mathbb{F}_q^{1\times N}$, $\bm{\beta}=(\beta_1,\beta_2,\cdots,\beta_N)\in\mathbb{F}_q^{1\times N}$, $\bm{f}=(f_1,f_2,\cdots,f_L)\in\mathbb{F}_q^{1\times L}$, all $\beta_i, i\in[N]$ are non-zero,  the terms $\alpha_1,\cdots, \alpha_N,f_1,\cdots,f_L$ are $L+N$ distinct elements of $\mathbb{F}_q$, $q\geq L+N$, $L\leq N/2$.  The sub-matrices $\mbox{GRS}^q_{N,\lfloor N/2\rfloor}(\bm{\alpha},\bm{\beta})$ and $\mbox{GRS}^q_{N,\lceil N/2\rceil}(\bm{\alpha},\bm{\beta})$ identified in \eqref{eq:MCV_Blocks}, are the generator matrices of $[N, \lfloor\frac{N}{2}\rfloor]$ and $[N, \lceil\frac{N}{2}\rceil]$ GRS codes, respectively (identical, when $N$ is even).
As a special case, setting $\beta_1=\cdots=\beta_N=1$, we have,
\begin{align}
\mbox{\rm CSA}^q_{N,L}(\bm{\alpha},\bm{f}) = \mbox{\rm QCSA}^q_{N,L}(\bm{\alpha},[1~~1\cdots~~1],\bm{f}).
\end{align}
\end{definition}

\begin{remark} 
$\mbox{QCSA}_{N,L}^{q}(\bm{\alpha},\bm{\beta},\bm{f})$ is the product of an $N\times N$ diagonal matrix $\mbox{Diag}(\beta_1,\beta_2,\cdots,\beta_N)$ and the $N\times N$ Cauchy-Vandermonde  matrix $\mbox{CSA}^q_{N,L}(\bm{\alpha},\bm{f})$ invoked by CSA schemes (cf. (11) of \cite{Jia_Jafar_MDSXSTPIR}). The former is full rank since $\beta_n\neq 0, \forall n\in [N]$, and the latter is full rank according to Lemma 1 of \cite{Jia_Jafar_MDSXSTPIR} because $\alpha_1,\cdots, \alpha_N,f_1,\cdots, f_N$ are distinct ($q\geq L+N$). Since multiplication with an invertible square matrix preserves rank, $\mbox{QCSA}^q_{N,L}(\bm{\alpha},\bm{\beta},\bm{f})$ is an invertible square matrix. 
\end{remark}

\section{Results}\label{sec:main}
Given a  CSA scheme over $\mathbb{F}_q$, we  translate it into a QCSA scheme over a QMAC by the following three steps.
\begin{enumerate}
\item From the CSA matrix $\mbox{CSA}_{N,L}^{q}(\bm{\alpha}, \bm{f})$, construct two QCSA matrices $\mbox{QCSA}_{N,L}^{q}(\bm{\alpha}, \bm{u}, \bm{f})\triangleq \bQ_N^{\bm{u}}$ and $\mbox{QCSA}_{N,L}^{q}(\bm{\alpha}, \bm{v}, \bm{f})\triangleq \bQ_N^{\bm{v}}$.
\item From $\bQ_N^{\bm{u}}, \bQ_N^{\bm{v}}$, construct a feasible $N$-sum Box, i.e., a MIMO MAC with channel matrix ${\bf M}_Q$.
\item Over the MIMO MAC, realize `over-the-air'  decoding of two instances of the CSA scheme. By the $N$-sum Box abstraction, this automatically maps to a quantum protocol (a QCSA scheme), and the efficiency gained by `over-the-air' decoding in the MIMO MAC translates into the superdense coding gain over the QMAC.
\end{enumerate}
These steps are explained next.

\subsection*{Step 1: $\mbox{\rm CSA}_{N,L}^{q}(\bm{\alpha}, \bm{f}) \rightarrow \mbox{\rm QCSA}_{N,L}^{q}(\bm{\alpha}, \bm{u}, \bm{f})\triangleq \bQ_N^{\bm{u}}$, and $\mbox{\rm QCSA}_{N,L}^{q}(\bm{\alpha}, \bm{v}, \bm{f})\triangleq \bQ_N^{\bm{v}}$}
Let $\bm{u} = [u_1 ~~ u_2 ~~ \cdots ~~ u_N] \in \Fq^{1\times N}$ and $\bm{v} = [v_1 ~~ v_2 ~~ \cdots ~~ v_N] \in \Fq^{1\times N}$. First, let $u_1, \cdots, u_N$ be $N$ arbitrary non-zero elements in $\Fq$. Then, we choose $v_j\in\mathbb{F}_q$ as
\begin{align}
    v_j&=\frac{1}{u_j}\left(\prod_{i\in[N],i\neq j }(\alpha_j-\alpha_i)\right)^{-1}&&\forall j\in[N],\label{eq:b_compute}
\end{align}
such that the sub-matrix $\mbox{\rm GRS}^q_{N,\lceil\frac{N}{2}\rceil}(\bm{\alpha},\bm{u})$ of $\mbox{\rm QCSA}_{N,L}^{q}(\bm{\alpha}, \bm{u}, \bm{f})\triangleq \bQ_N^{\bm{u}}$ and the sub-matrix $\mbox{\rm GRS}^q_{N,\lfloor\frac{N}{2}\rfloor}(\bm{\alpha},\bm{v})$ of $\mbox{\rm QCSA}_{N,L}^{q}(\bm{\alpha}, \bm{v}, \bm{f})\triangleq \bQ_N^{\bm{v}}$ satisfy
\begin{align}
    \Big(\mbox{\rm GRS}^q_{N,\lceil\frac{N}{2}\rceil}(\bm{\alpha},\bm{u})\Big)^\top\mbox{\rm GRS}^q_{N,\lfloor\frac{N}{2}\rfloor}(\bm{\alpha},\bm{v})=\bm{0}_{\lceil\frac{N}{2}\rceil \times \lfloor\frac{N}{2}\rfloor},\label{eq:dualsub}
\end{align}
according to \eqref{def:v} and \eqref{eq:guarantee}. As compact notation for the resulting QCSA matrices, let us define $\bQ_N^{\bm{u}}\triangleq\mbox{QCSA}_{N,L}^{q}(\bm{\alpha}, \bm{u}, \bm{f})$ and $\bQ_N^{\bm{v}}\triangleq\mbox{QCSA}_{N,L}^{q}(\bm{\alpha}, \bm{v}, \bm{f})$.

\subsection*{Step 2: A suitable $N$-Sum Box $({\bf M}_Q)$}
The $N$-sum box is specified by the following theorem.
\begin{theorem}\label{thm:QCSA}
For the $\bQ_N^{\bm{u}}$ and $\bQ_N^{\bm{v}}$ constructed in Step 1, there exists a feasible $N$-sum box $\by=\bM_Q\bx$ in $\mathbb{F}_q$ with the $N\times 2N$ channel matrix,
\begin{align}
&\scalebox{0.81}{$\bM_Q=$}
\scalebox{0.7}{$\left[\begin{array}{ccc|ccc}
\bI_L&\mathbf{0}_{L\times \lceil N/2\rceil}&\mathbf{0}&\mathbf{0}&\mathbf{0}&\mathbf{0}\\
\mathbf{0}&\mathbf{0}&\bI_{\lfloor N/2\rfloor -L}&\mathbf{0}&\mathbf{0}&\mathbf{0}\\
\mathbf{0}&\mathbf{0}&\mathbf{0}&\bI_L&\mathbf{0}_{L\times \lfloor N/2\rfloor }&\mathbf{0}\\
\mathbf{0}&\mathbf{0}&\mathbf{0}&\mathbf{0}&\mathbf{0}&\bI_{\lceil N/2\rceil -L}
\end{array}\right]
$}
\scalebox{0.7}{$\left[\begin{array}{cc}
\bQ_N^{\bm{u}}&\mathbf{0}\\
\mathbf{0}&\bQ_N^{\bm{v}}
\end{array}\right]^{-1}
$}\label{eq:thmQCSA}
\end{align}
\end{theorem}
A proof of Theorem \ref{thm:QCSA} is presented in Section \ref{sec:proofQCSA}. 

\subsection*{Step 3: QCSA Scheme as `Over-the-Air' CSA}\label{sec:csa2qcsa}
Using the MIMO MAC with channel matrix ${\bf M}_Q$ identified in Step $2$, we now describe how to achieve `over-the-air' decoding of several instances of CSA schemes. 

First consider the case where we are given a CSA scheme with $L\leq N/2$. With $2$ instances of the CSA scheme, we have,
\begin{align*}
\scalebox{0.8}{$\begin{bmatrix}
\bm{A}(1)\\
\bm{A}(2)
\end{bmatrix}$}&=
\scalebox{0.8}{$\begin{bmatrix}
\mbox{CSA}^q_{N,L}(\bm{\alpha},\bm{f})&\mathbf{0}\\
\mathbf{0}&\mbox{CSA}^q_{N,L}(\bm{\alpha},\bm{f})
\end{bmatrix}
\begin{bmatrix}
\bm{X}_{\delta\nu}(1)\\
\bm{X}_{\delta\nu}(2)
\end{bmatrix}$}
\end{align*}
which retrieves $2L$ desired symbols at the download cost of $2N$ symbols. Now the corresponding QCSA scheme (over-the-air MIMO MAC) is obtained as follows.
\begin{align}
&\underbrace{\begin{bmatrix}
y_1\\
\vdots\\
y_N
\end{bmatrix}}_{\by}=\bM_Q
\underbrace{\begin{bmatrix}
\mbox{Diag}(\bm{u})&\mathbf{0}\\
\mathbf{0}&\mbox{Diag}(\bm{v})
\end{bmatrix}
\begin{bmatrix}
\bm{A}(1)\\
\bm{A}(2)
\end{bmatrix}}_{\bx}\label{eq:serverscale}\\
&=\bM_Q\begin{bmatrix}
\mbox{Diag}(\bm{u})&\mathbf{0}\\
\mathbf{0}&\mbox{Diag}(\bm{v})
\end{bmatrix}\notag\\
&~~~\cdot\begin{bmatrix}
\mbox{CSA}^q_{N,L}(\bm{\alpha},\bm{f})&\mathbf{0}\\
\mathbf{0}&\mbox{CSA}^q_{N,L}(\bm{\alpha},\bm{f})
\end{bmatrix}
\begin{bmatrix}
\bm{X}_{\delta\nu}(1)\\
\bm{X}_{\delta\nu}(2)
\end{bmatrix}\\
&=\bM_Q\begin{bmatrix}
\bQ_N^{\bm{u}}&\mathbf{0}\\
\mathbf{0}&\bQ_N^{\bm{v}}
\end{bmatrix}\begin{bmatrix}
\bm{X}_{\delta\nu}(1)\\
\bm{X}_{\delta\nu}(2)
\end{bmatrix}\\
&=
\scalebox{0.75}{$\left[\begin{array}{ccc|ccc}
\bI_L&\mathbf{0}_{L\times \lceil N/2\rceil}&\mathbf{0}&\mathbf{0}&\mathbf{0}&\mathbf{0}\\
\mathbf{0}&\mathbf{0}&\bI_{\lfloor N/2\rfloor -L}&\mathbf{0}&\mathbf{0}&\mathbf{0}\\
\mathbf{0}&\mathbf{0}&\mathbf{0}&\bI_L&\mathbf{0}_{L\times \lfloor N/2\rfloor }&\mathbf{0}\\
\mathbf{0}&\mathbf{0}&\mathbf{0}&\mathbf{0}&\mathbf{0}&\bI_{\lceil N/2\rceil -L}
\end{array}\right]$}
\scalebox{0.81}{$\begin{bmatrix}
\bm{X}_{\delta\nu}(1)\\
\bm{X}_{\delta\nu}(2)
\end{bmatrix}$}\notag\\
&=\begin{bmatrix}\delta_1(1),\cdots,\delta_L(1),\bm{\nu}_{(\leftharpoondown)}(1), \delta_1(2),\cdots,\delta_L(2),\bm{\nu}_{(\leftharpoondown)}(2)\end{bmatrix}^\top\notag
\end{align}
where the $N$ entries of $\bm{u}$ are non-zero, and $\bm{v}$ is specified in \eqref{eq:b_compute} and we used the compact notation for the interference symbols, i.e., $\bm{\nu}_{(\leftharpoondown)}(1)$ to represent the last $\lfloor N/2\rfloor -L$ symbols of the vector $\bm{\nu}(1)=(\nu_1(1),\cdots,\nu_{N-L}(1))$, and $\bm{\nu}_{(\leftharpoondown)}(2)$ to represent the last $\lceil N/2\rceil -L$ symbols of the vector $\bm{\nu}(2)=(\nu_1(2),\cdots,\nu_{N-L}(2))$.  Multiplication by $\mbox{Block-Diag}(\mbox{Diag}(\bm{u}), \mbox{Diag}(\bm{v}))$ in \eqref{eq:serverscale} simply involves each Server $j, j\in[N]$ scaling its answers of the two instances of the CSA scheme $A_j(1), A_j(2)$ by $u_j, v_j$ respectively and applying $u_j A_j(1), v_j A_j(2)$ to the inputs of the $N$-sum box (MIMO MAC) corresponding to that server. Evidently, all $2L$ desired symbols are recovered, and the total download cost is $N$ qudits (one qudit from each server), for a normalized download cost of $N/(2L)$ qudits per desired dit. The improvement from $N/L$ (classical CSA) to $N/2L$ (QCSA) reflects the factor of $2$ superdense coding gain in communication efficiency.

If $L > N/2$, we  discard  `redundant' servers (see Remark \ref{rem:redundant}) and only employ $N^\prime =2N-2L < N$ servers, choosing a CSA scheme with $L' =  N^\prime / 2$, such that $N-L=N'-L'$, i.e., the dimensions of interference are preserved, which results in a download cost of $N'/(2L')=1$ qudit per desired symbol.

\section{Proof of Theorem \ref{thm:QCSA}}\label{sec:proofQCSA}
For a permutation $\pi:[2N]\rightarrow[2N]$, define,
    \begin{align}
        \bP_{\pi} \triangleq 
        \begin{bmatrix}
            \be_{\pi(1)} & \be_{\pi(2)} & \cdots & \be_{\pi(2N)}
        \end{bmatrix},
    \end{align}
where $\be_{i}$ is the $i^{th}$ column of $\bI_{2N}$. 
When $\bP_{\pi}$ is multiplied to the right of a $2N \times 2N$ matrix $\bA$, the columns of $\bA$ will be permuted according to the permutation $\pi$. Also, note that $\bP_{\pi}^{-1} = \bP_{\pi^{-1}} = \bP_{\pi}^\top$, 
where $\pi^{-1}$ is the inverse permutation of $\pi$.

Now let us prove Theorem \ref{thm:QCSA} by specifying the $N$-sum box construction according to \eqref{eq:feasible}. Following the CSS construction \cite{CSS_CS,CSS_S}, i.e., the fact that placing the generator matrices of two codes that are dual to each other on the diagonal creates an SSO matrix, let us define,
\begin{align}
\tilde{\bG}&=\begin{bmatrix}\mbox{GRS}^q_{N,\lceil N/2\rceil}(\bm{\alpha},\bm{u})&\mathbf{0}\\ 
\mathbf{0}&\mbox{GRS}^q_{N,\lceil N/2\rceil}(\bm{\alpha},\bm{v})\end{bmatrix},
\end{align}
where $\mbox{GRS}^q_{N,\lceil N/2\rceil}(\bm{\alpha},\bm{u})$ and $\mbox{GRS}^q_{N,\lceil N/2\rceil}(\bm{\alpha},\bm{v})$ are sub-matrices of $\bQ^{\bm u}_N$ and $\bQ^{\bm v}_N$, respectively.
If $N$ is even, then $\tilde{\bG}$ is a $2N\times N$  square matrix, and we will choose $\bG=\tilde{\bG}$ for the $N$-sum Box construction as in \eqref{eq:feasible}. If $N$ is odd,  then $\tilde{\bG}$ is a $2N\times (N+1)$ matrix, i.e., it has an extra column. In this case we will choose $\bG$ as the $2N\times N$ left-sub-matrix of $\tilde{\bG}$, leaving out the $(N+1)^{th}$ column. Thus, 
\begin{align}
\bG&=\tilde{\bG}(:,1:N)\\
&=\begin{bmatrix}\mbox{GRS}^q_{N,\lceil N/2\rceil}(\bm{\alpha},\bm{u})&\mathbf{0}\\ 
\mathbf{0}&\mbox{GRS}^q_{N,\lfloor N/2\rfloor}(\bm{\alpha},\bm{v})\end{bmatrix}.\label{eq:Gform}
\end{align}
Note that ${\bf G}$ is a sub-matrix of $\mbox{Block-Diag}(\bQ_N^{\bm u}, \bQ_N^{\bm v}) = \begin{bmatrix}\bQ_N^{\bm u} & \bzero \\ \bzero & \bQ_N^{\bm v}\end{bmatrix}$. Let  $\bH$ in \eqref{eq:feasible} be chosen as the remaining columns of $\mbox{Block-Diag}(\bQ_N^{\bm u}, \bQ_N^{\bm v})$ after eliminating the columns that are present in $\bG$. Thus,
\begin{align}
&\left[\begin{array}{c|c}
\bG & \bH
\end{array}\right]_{2N\times 2N}
=\notag\\
&{\scriptsize\left[\begin{array}{c|c;{2pt/2pt}c;{2pt/2pt}c;{2pt/2pt}c}
\tilde{\bG}(:,1:N) & \begin{matrix}\tilde{\bH}^{\bm u}\\\mathbf{0} \end{matrix}& \begin{matrix}\mathbf{0}\\ \tilde{\bH}^{\bm v}(:,1:L)\end{matrix} & {\color{black!50}\tilde{\bG}(:,N+1)} & \begin{matrix}\mathbf{0}\\ \tilde{\bH}^{\bm v}(:,L+1:\lfloor\frac{N}{2}\rfloor)\end{matrix} 
\end{array}\right]}
\label{eq:HGform}\\
&=
\begin{bmatrix}
    \bQ_N^{\bm u} & \mathbf{0}\\
    \mathbf{0} & \bQ_N^{\bm v}
\end{bmatrix}
\bP_{\pi},
\end{align}
The column $\tilde{\bG}(:,N+1)$ does not appear in \eqref{eq:HGform} (i.e., it is  empty) if $N$ is even, because in that case $\tilde{\bG}$ has only $N$ columns. $\tilde{\bH}^{\bm u} = \begin{bmatrix}\tilde{\bH}^{\bm u}(:,1:L) & \tilde{\bH}^{\bm u}(:,L+1:\lfloor\frac{N}{2}\rfloor)\end{bmatrix}$, $\tilde{\bH}^{\bm v}(:,1:L)$ and $\tilde{\bH}^{\bm v}(:,L+1:\lfloor\frac{N}{2}\rfloor)$ are specified in \eqref{eq:MCV_Blocks}. Evidently, the $2N\times 2N$ matrix $\begin{bmatrix}\bG&\bH\end{bmatrix}$ has full rank $2N$ as required for an $N$-sum Box construction \eqref{eq:feasible}, because $\bQ_N^{\bm u}, \bQ_N^{\bm v}$ have full rank $N$,  the rank of a block-diagonal matrix is the sum of the ranks of its blocks so $\mbox{Block-Diag}(\bQ_N^{\bm u}, \bQ_N^{\bm v})$ has rank $2N$, and  $\bP_\pi$ is an invertible square matrix which preserves rank. The permutation $\pi$ is explicitly expressed as,
\begin{align}
\pi &=(\pi(1),\cdots,\pi(2N))\\
&=\Big(L+1, L+2, \cdots, L+\lceil N/2\rceil, \notag\\
&~~~N+L+1, N+L+2, \cdots,N+L+\lfloor N/2\rfloor\notag\\
&~~~1,2,\cdots, L, ~~~L+\lceil N/2\rceil+1, L+\lceil N/2\rceil+2, \cdots, N,\notag\\
&~~~N+1, N+2, \cdots, N+L,\label{eq:HGperm}\\
&~~~N+L+\lfloor N/2\rfloor +1,N+L+\lfloor N/2\rfloor +2, \cdots, 2N\Big).\notag
\end{align}
Note that $\bP_{\pi}$ moves the columns of $\mbox{Block-Diag}(\bQ_N^{\bm u}, \bQ_N^{\bm v})$, with indices in $\bigg[L+1: L+\lceil {N}/{2} \rceil\bigg] \cup \bigg[N+L+1: N+L+\lfloor{N}/{2}\rfloor\bigg]$, to the left-most part.

To establish the feasibility of this $N$-sum Box as in \eqref{eq:feasible},  it remains to prove that $\bG$ is SSO (see definition in Footnote \ref{fn:sso}, and originally in \cite{Allaix_N_sum_box}). Since $\begin{bmatrix}\bG&\bH\end{bmatrix}$ has full rank $2N$, it follows that $\bG$ has rank $N$. Furthermore, the SSO property follows from the CSS construction \cite{CSS_CS,CSS_S}, also easily verified next in \eqref{eq:ssoQCSA1}-\eqref{eq:ssoQCSA3}, where we use the compact notation,
\begin{align}
&\overline\Gamma\triangleq \mbox{GRS}^q_{N,\lceil N/2\rceil}(\bm{\alpha},\bm{u}),~~\underline\Gamma \triangleq \mbox{GRS}^q_{N,\lfloor N/2\rfloor}(\bm{\alpha},\bm{v}).\\
&\bG^\top \bJ \bG = \begin{bmatrix}\overline\Gamma^\top&\mathbf{0}\\ 
\mathbf{0}&\underline\Gamma^\top\end{bmatrix}
\begin{bmatrix}
    \mathbf{0} & -\bI_{N}\\
    \bI_{N} & \mathbf{0}
\end{bmatrix}
\begin{bmatrix}\overline\Gamma&\mathbf{0}\\ 
\mathbf{0}&\underline\Gamma\end{bmatrix}\label{eq:ssoQCSA1}\\
&= \begin{bmatrix}\overline\Gamma^\top&\mathbf{0}\\ 
\mathbf{0}&\underline\Gamma^\top\end{bmatrix}
\begin{bmatrix}
\mathbf{0}&-\underline\Gamma\\
\overline\Gamma&\mathbf{0}
\end{bmatrix}\stackrel{\eqref{eq:dualsub}}{=} \bzero_{N\times N}\label{eq:ssoQCSA3}
\end{align}

\noindent  Therefore,  this $N$-sum box is feasible according to \eqref{eq:feasible}.

For the specified $\bG, \bH$ matrices, the transfer matrix $\bM_{Q}$ is,
\begin{align}
    \bM_{Q}&= 
    \begin{bmatrix}
        \mathbf{0}_N & \bI_N
    \end{bmatrix}
    \begin{bmatrix}
        \bG & \bH
    \end{bmatrix}^{-1}\\
    &= 
    \begin{bmatrix}
        \mathbf{0}_N & \bI_N
    \end{bmatrix}
    \mathbf{P}_{\pi}^{-1}
    \begin{bmatrix}
        \bQ_N^{\bm u} & \mathbf{0}\\
        \mathbf{0} & \bQ_N^{\bm v}
    \end{bmatrix}^{-1}\label{eq:verify}
\end{align}
Recall that $\bP_{\pi}^{-1} = \bP_{\pi^{-1}}$. 
Now it is easily verified that $[\mathbf{0}_N~~\mathbf{I}_{N}]\mathbf{P}_{\pi}^{-1}$ in \eqref{eq:verify} is  the same as the  matrix in \eqref{eq:thmQCSA}.$\hfill\square$

\section{Conclusion}\label{sec:con}
Using the $N$-sum box abstraction  \cite{Allaix_N_sum_box}, this work translated existing CSA coding schemes  for classical MAC settings into Quantum CSA coding schemes, for entanglement-assisted QMACs. This leads to immediate applications to Quantum PIR with secure and MDS-coded storage (Q-MDS-XSTPIR), as well as Quantum SDBMM. In both cases, the rate achieved with the QCSA scheme can be expressed as $R^Q=\min\{1,2R^{C}\}$, where $R^C$ is the rate achieved by the  CSA scheme in the corresponding classical setting. Recent results in QPIR, QTPIR, QMDSTPIR can be recovered as  special cases. An important direction for future work is to explore settings where certain qudits are erased or lost, which would extend the scope of QCSA schemes to allow stragglers.

\section*{Acknowledgment}
The authors gratefully acknowledge helpful discussions with M. Allaix from Aalto University and Y. Yao from UC Irvine.

\bibliographystyle{IEEEtran}
\bibliography{thesis}
\end{document}